\documentclass[traditabstract]{aa}
\usepackage{txfonts}
\usepackage{natbib}
\usepackage{graphicx}
\usepackage{rotating}
\usepackage{float}
\newcommand{\Msun}{M$_{\odot}$}

\newcommand{\Lsun}{L$_{\odot}$}

\newcommand{\Rsun}{R$_{\odot}$}

\begin{document}

\title{The Near-Infrared Outflow and Cavity of the proto-Brown Dwarf Candidate ISO-Oph 200\thanks{Based on Observations collected with SINFONI at the Very Large Telescope on Cerro Paranal (Chile), 
operated by the European Southern Observatory (ESO). Program ID: 097.C-0732(A).}}

\author{Whelan, E.T. \inst{1}
  \and 
  Riaz, B. \inst{2}
   \and 
   Rouz{\'e}, B. \inst{1}}

\institute{Maynooth University Department of Experimental Physics, National University of Ireland Maynooth, Maynooth Co. Kildare, Ireland
 \and 
 Max-Planck-Institut für Extraterrestrische Physik, Giessenbachstrasse 1, D-85748 Garching, Germany
}

\titlerunning{The ISO-Oph 200 NIR Outflow} 
\date{}

\abstract {In this letter a near-infrared integral field study of a proto-brown dwarf candidate is presented. A $\sim$ 0\farcs5 blue-shifted outflow is detected in both H$_{2}$ and [Fe II] lines at V$_{sys}$ = (-35~$\pm$ 2)~km~s$^{-1}$ and V$_{sys}$ = (-51~$\pm$ 5)~km~s$^{-1}$ respectively. In addition, slower ($\sim$ $\pm$~10~km~s$^{-1}$) H$_{2}$ emission is detected out to $<$ 5\farcs4, in the direction of both the blue and red-shifted outflow lobes but along a different position angle to the more compact faster emission. It is argued that the more compact emission is a jet and the extended H$_{2}$ emission is tracing a cavity. The source extinction is estimated at A$_{v}$ = 18 mag $\pm$ 1 mag and the outflow extinction at Av = 9 mag $\pm$ 0.4 mag. 
The H$_{2}$ outflow temperature is calculated to be 1422 K $\pm$~255~K and the electron density of the [Fe II] outflow is measured at $\sim$ 10000 cm$^{-3}$. Furthermore, the mass outflow rate is estimated at $\dot{M}_{out[H2]}$ = 3.8 $\times$ 10$^{-10}$~\Msun~yr$^{-1}$ and $\dot{M}_{out[FeII]}$ = 1 $\times$ 10$^{-8}$~\Msun~yr$^{-1}$. $\dot{M}_{out[FeII]}$ takes a Fe depletion of $\sim$ 88$\%$ into account. The depletion is investigated using the ratio of the [FeII]~1.257~$\mu$m and [PII]~1.188~$\mu$m lines. Using the Pa$\beta$ and Br$\gamma$ lines and a range in stellar mass and radius $\dot{M}_{acc}$ is calculated to be (3 - 10) $\times$ 10$^{-8}$~\Msun~yr$^{-1}$. Comparing these rates puts the jet efficiency in line with predictions of magneto-centrifugal models of jet launching in low mass protostars. This is a further case of a brown dwarf outflow exhibiting analogous properties to protostellar jets.}

\maketitle


\section{Introduction}

{Outflows play an important role in the star formation process
and are driven by young stellar objects (YSOs) in the Class 0 to
Class II evolutionary stages \citep{Frank14}. Thus, they can
be observed across a large range in wavelength \citep{Whelan14a}.
Also, as different wavelength regimes can offer complementary observational constraints to launching models, multi-wavelength studies have proven important \citep{Nisini16}. Near-infrared (NIR) integral field observations are a particularly
useful tool for studies of jet launching in Class 0/I sources which
are still very much embedded in their natal clouds \citep{Davis11}. The J,H, and K NIR bands cover Fe II forbidden and H$_{2}$ ro-vibrational emission lines that are strong tracers of jets. These lines can be used to infer jet properties such as temperature and density and to measure the extinction towards the jet \citep{GLopez2013}. Integral field spectroscopy provides spectro-images which allow the properties of the jets to be mapped as a function of velocity and distance \citep{Davis11}. Such studies have found the [Fe II] and H$_{2}$ lines to be tracing different components of the jet, with the forbidden emission associated with faster, hot dense material and the molecular emission tracing lower excitation, slower molecular gas \citep{Caratti2006, GLopez2010}. Additionally, H$_{2}$ is often found to trace cavities which are postulated to be evacuated by a wide-angled wind \citep{Davis11}.

Recently, we reported the first detection of a large-scale Herbig-Haro optical jet driven by a Class I proto-brown dwarf (proto-BD) that shows several of the well-known features seen in protostellar HH jets \citep{Riaz17}. Here we present the results of a near-infrared study of an outflow from a proto-BD candidate, which also shows characteristics similar to protostellar jets. The driving source of the outflow is a YSO named ISO-Oph 200 ($\alpha$ = 16:31:43.8, $\delta$ = - 24:55:24.5). The near- to mid-infrared (2-24~$\mu$m) spectral slope of the SED for ISO-Oph 200 is consistent with a Class I classification \citep{Evans2009}. We have conducted physical and chemical modelling of sub-millimeter/millimeter continuum and molecular line data for this object, and find the results to be consistent with an early Stage 0+I evolutionary stage (Riaz et al. in prep). The total (dust + gas) mass for ISO-Oph 200 as derived from the sub-millimeter 850~$\mu$m flux is 0.06$\pm$0.01~\Msun\ with a bolometric luminosity of $\sim$ 0.09$\pm$0.02~\Lsun. The mass of the central object in this Class I system can be constrained to within the sub-stellar mass regime using the measured bolometric luminosity and numerical simulations of stellar evolution, as discussed in \cite{Riaz2016}. Considering the very low mass reservoir in the (envelope+disk) for this system, and the presence of an outflow that will further dissipate the envelope material, ISO Oph-200 will likely have a final mass below the sub-stellar limit, and can be considered as a strong candidate proto-BD.



\begin{figure*}
\begin{center}
   \includegraphics[width=13cm, trim= 0cm 0cm 0cm 0cm, clip=true]{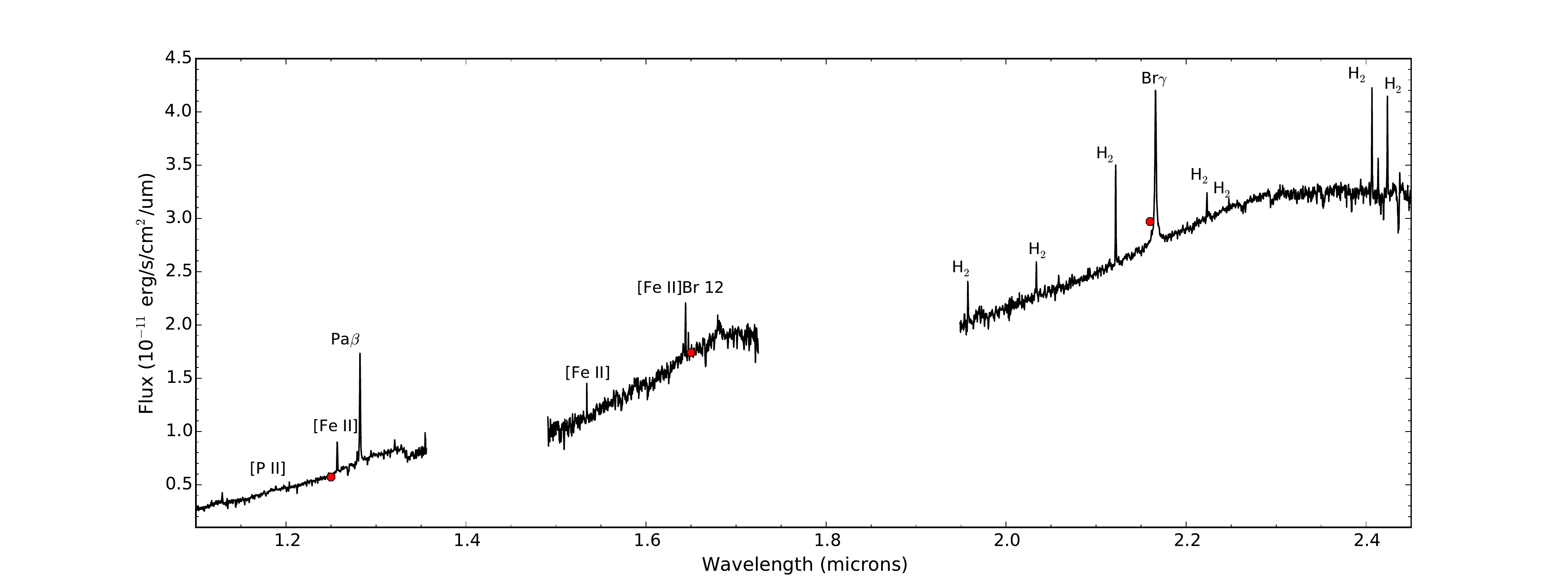}
     \caption{The flux calibrated, telluric subtracted, J, H, and K spectra of ISO-Oph 200. Spectra were extracted from a 2~\arcsec $\times$ 2~\arcsec region centred on the source. The red dots are the 2MASS magnitudes of the source. The fluxes of the marked emission lines are given in Table 1.}
  \label{spec_full}
  \end{center}     
\end{figure*}


\section{Observations and data reduction}

ISO-Oph 200 was observed at the Very Large Telescope (VLT) using the integral field spectrograph SINFONI, on April 19 2016 (H and K band, seeing = 1.1), May 12 2016 (J band, seeing = 1\farcs3), and May 13 2016 (K band, seeing = 0\farcs55). SINFONI is
a medium resolution instrument with R =2000, 3000, 4000, in the J, H and K bands respectively \citep{Gillessen05}. Observations were made with the 8~\arcsec $\times$ 8~\arcsec field of view (FOV) and therefore a pixel scale of 0\farcs125. The recommended observing mode of ABB$^{'}$A$^{'}$ sky nodding and a jittering between the source and sky positions was chosen. The total exposure times for the J, H and K bands were 2250~s, 240~s and 120~s (both observations) respectively. The SINFONI pipeline was used to produce the wavelength calibrated, sky subtracted datacubes. The wave- length calibration provided by the pipeline was checked using the OH lines and an average shift of 40 km s$^{-1}$ was measured across all four cubes. This was comparable to the systematic wavelength shift of 2.2~\AA\ reported by \cite{GLopez2013}. The telluric subtraction and flux calibration were carried out in the usual way using standard star observations \citep{GLopez2013}. The extracted J, H and K band flux calibrated spectra are presented in Figure 1 with the J, H and K magnitudes as measured by the Two Micron All Sky Survey (2MASS) over plotted \citep{Cutri03}.

\section{Results}
The fluxes of the lines marked in Figure 1 are given in Table \ref{table}. Numerous [Fe II] and H$_{2}$ lines are detected along with the H~I lines Pa , Br  and Br 12, which are known to primarily trace accretion. In Figure 2, spectro-images in the H$_{2}$ 1-0S(1) and [Fe II]~1.257~$\mu$m lines are presented.
A compact blue-shifted outflow is detected in all lines, while larger scale, slower, red and blue-shifted extended emission is detected in H$_{2}$ only. Evidence is presented below that this slower more extended emission is a cavity.

\subsection{Extinction and mass accretion rate estimates}
The source visual extinction can firstly be derived using the J, H and K magnitudes. Following the method as described in \cite{Davis11}, where the source is plotted on a NIR color - color diagram and its position compared to the relevant locus, the source extinction is estimated at A$_{v}$ = 12 $\pm$ 1~mag. A second approach is to estimate A$_{v}$ simultaneously with L$_{acc}$ by assuming that L$_{acc}$ measured from Br$\gamma$ and Pa$\beta$ agree (see the following paragraph of this section for a description of how L$_{acc}$ is calculated. This gives an estimate of A$_{v}$ = 18~$\pm$~1~mag. Thirdly one can use the outflow lines extracted at the source position to derive the outflow extinction at this point. This can be taken as a lower limit to the source extinction. Forbidden transitions of [Fe II] that arise from the same upper level are used to estimate A$_{v}$ with the [Fe II] 1.644~$\mu$m / 1.25~$\mu$m ratio frequently used \citep{Nisini16}. From this ratio the outflow extinction here is estimated at 9.0~$\pm$~0.4~mag. The Einstein coefficients as reported in \cite{Giannini2015} were adopted along with the extinction law of \cite{Cardelli89}. Considering that the estimate of A$_{v}$ from the ratio of the [Fe II]~1.644~$\mu$m / 1.25~$\mu$m lines, is consistent with the jet extinction being a lower limit to the source extinction, the estimate of A$_{v}$ 18~$\pm$~1~mag is adopted for the source.


The mass accretion rate ($\dot{M}_{acc}$) was estimated from the luminosity of the Pa$\beta$ and Br$\gamma$ lines. The procedure as outlined in \cite{Whelan14b} was followed along with the L$_{line}$ and L$_{acc}$ relations of \cite{Alcala17}. Fluxes were corrected for a source extinction of A$_{v}$ = 18~mag. For the mass, a range of 0.06~\Msun\ to 0.1~\Msun\ was adopted with corresponding radii of 0.9~\Rsun\ and 0.5~\Rsun\ respectively. This range is based on estimates of the source luminosity and the cold and hybrid accretion models of \cite{Baraffe17}. The average value of log($\dot{M}_{acc}$) is found to be in the range -7.0~$\pm$~0.4 to -7.5~$\pm$~0.4. \cite{GLopez2013} found log($\dot{M}_{acc}$) $\sim$ -6.5 for the very low mass protostar IRS54 (using the Br$\gamma$ line) therefore the results presented here are consistent with the idea that ISO-Oph 200 is a proto-BD. In \cite{Riaz17} and \cite{Riaz15} we investigated $\dot{M}_{acc}$ for the Class I BDs Mayrit 1701117 and Mayrit 1082188, and report values in the range log($\dot{M}_{acc}$) = -9.3 to -9. These values were calculated using optical spectra and are likely underestimated. This is due to scattering effects which are more prominent in this wavelength range and the likely obscuration of some component of the accretion emission. Here the Pa$\beta$ and Br$\gamma$ emission lines regions were analysed using spectro-astrometry to check for any contribution from the outflow and none was found \citep{Whelan2004}.

\subsection{Outflow morphology, kinematics and physical properties}
In Figure 2, continuum subtracted H$_{2}$ 1-0S(1) and [Fe II]~1.257~$\mu$m spectro-images of the source in different velocity bins are presented. All velocities are systemic and the velocity bins were chosen to highlight the outflow features, as outlined in the figure caption. In Table 1 the properties of the outflow features are compared. In the top panels of Figure 2 the emission in a predominantly blue-shifted velocity range is shown. The peak velocity of the H$_{2}$ and [Fe II] emission is measured at (-35~$\pm$~2)~km s$^{-1}$ and (-51~$\pm$~5)~km s$^{-1}$ respectively. These radial velocities are in line with the radial velocities of Class II BD outflows \citep{Whelan14a}. A 2D fitting of the emission regions shows that the H$_{2}$ emission peaks at a distance of 0\farcs7 from the source along a position angle (PA) of 230$^{\circ}$ and the [Fe II] emission at a distance of 0\farcs5 from the source along a position angle (PA) of 215$^{\circ}$. The emission in this velocity bin is compatible with a small-scale jet where the jet width is not spatially resolved in the observations. The difference between the velocity of the H$_{2}$ and [Fe II] jets is consistent with the fact that the H$_{2}$ tracers the outer layers of the jet where material is cooler and slower \citep{Davis11}. 

 \begin{figure}
\begin{center}
   \includegraphics[width=10cm,trim= 2cm 0cm 0cm 0cm, clip=true]{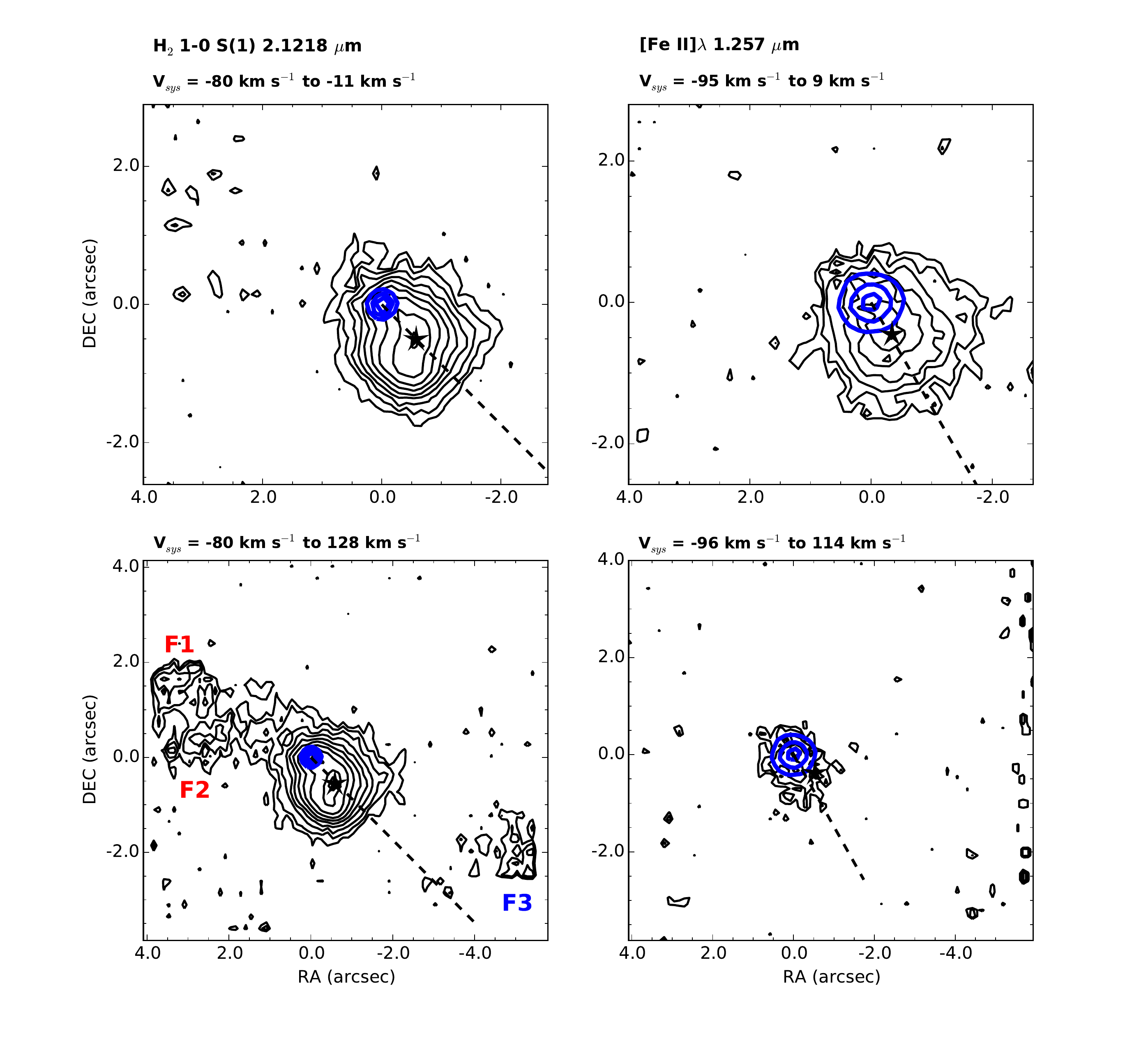}
     \caption{Continuum subtracted spectro-images of the ISO-Oph 200 outflow in the H$_{2}$ 1-0 S(1) and [Fe II] 1.257~$\mu$m lines. Velocities have been corrected for the systemic velocity of the source. {\bf Top panels:} Here the emission in a mainly blue-shifted velocity range is shown to highlight the blue-shifted jet emission only. Black contours begin at 3$\sigma$ and increase in multiples of 1.5. 1$\sigma$ = 4.5 $\times$ 10$^{-15}$ erg cm$^{-2}$ s$^{-1}$ $\mu$m$^{-1}$. The blue contours show the continuum emission and they correspond to  50$\%$, 75$\%$ and 95$\%$ of the peak flux of the continuum. The black stars marks the position of the emission peaks and the emission PA is delineated by the dashed line). {\bf Bottom panels:} Here the emission in the velocity ranges -80 km s$^{-1}$ to 128 km s$^{-1}$ (H$_{2}$) and -96 km s$^{-1}$ to 114 km s$^{-1}$ ([Fe II]) is shown. This range was chosen to show simultaneously the extended blue and red-shifted low velocity H$_{2}$ emission which it is argued traces the cavity. The chosen range also shows that there is no red-shifted counterpart to the blue-shifted jet emission and that the cavity emission is not seen in [FeII]. Contours start at 3$\sigma$ and increase in multiples of 1.5. 1$\sigma$ = 5 $\times$ 10$^{-15}$ erg cm$^{-2}$ s$^{-1}$ $\mu$m$^{-1}$. The continuum emission is plotted in the same way as for the top panels.} 
  \label{channelmaps1} 
  \end{center}     
\end{figure}

In the bottom panels the H$_{2}$ and [Fe II] emission in the velocity range -96 km s$^{-1}$ to 128 km s$^{-1}$ is shown. The velocity range was extended to 128 km s$^{-1}$ to check for a red-shifted jet lobe along the same PA and with a similar velocity to the blue-shifted lobe. None is detected. However, low velocity ($\sim$ 10 km s$^{-1}$), fainter (also see Figure \ref{cavity}), red and blue-shifted spatially resolved H$_{2}$ emission is seen at distance of 2\farcs6 to 5\farcs4 and along a different PA to the faster emission. The main emission features here are marked F1, F2 and F3 in Figure 2. The velocity, morphology and PA of this emission would suggest that it is tracing a cavity around the collimated jet. This emission is not detected in [Fe II] and is discussed further in Section 4.

\begin{table}
\centering
\begin{tabular}{ccccc}        
\hline \hline 
Outflow &V$_{sys}$    &Offset  &PA 
\\
Feature &(\small{km~s$^{-1}$})  &(\arcsec) &($\circ$) 
\\
\hline
H$_{2}$ Jet &-35~$\pm$2  &0\farcs7 &230 
\\
Fe II Jet &-51~$\pm$~5  & 0\farcs5 &215 
\\
F1   &4~$\pm$~3  &4\farcs0  &67 
\\ 
F2   &10~$\pm$~3  &2\farcs6 &80 
\\
F3  &-14~$\pm$~5  &5\farcs4 &248 
\\
\hline 
\end{tabular}
\caption{Outflow emission features and corresponding properties. Offsets are with respect to the source position and along the given PA. PAs are measured E of N. It is argued that F1, F2 and F3 form part of a cavity around the jet.}
\label{tableoutflow}
\end{table}

A H$_{2}$ excitation diagram can be constructed by plotting the natural log of the column densities (divided by the statistical weight), of a number of H$_{2}$ lines against the upper energy level of the line. This can be used to derive the temperature and to constrain the extinction of the line emitting region \citep{Davis11}. In a thermalised gas at a single temperature the points will lie along a straight line, the slope of which is the temperature \citep{Caratti2006}. The extinction is constrained by investigating the scatter of the points about this straight line. Correcting the line fluxes for extinction should reduce the scatter of the points about a linear fit. Therefore, varying A$_{v}$ and analysing the goodness of the fit leads to an estimate of A$_{v}$ \citep{Todd06}. A H$_{2}$ excitation diagram for the ISO-Oph 200 jet was constructed (see Figure \ref{rot}) using the fluxes given in Table 1. For A$_{v}$ = 9~mag, T = 1442~$\pm$~255~K is measured. This result is in agreement with the temperatures reported by \cite{GLopez2013} for the very low mass protostar IRS54. By varying the extinction it is found that the analysis of the excitation diagrams is in a agreement with the A$_{v}$ estimates presented in Section 3.1. 

Using the ratios of the [Fe II] lines and the models of \cite{Takami06}, the electron density (n$_{e}$) can be estimated. Studies have found that the 1.600~$\mu$m / 1.644~$\mu$m, 1.664~$\mu$m / 1.644~$\mu$m, and 1.677~$\mu$m / 1.644~$\mu$m ratios are more accurate tracers of the density than the 1.534~$\mu$m / 1.644~$\mu$m ratio. However, due to the non-detection of the 1.600~$\mu$m, 1.664~$\mu$m and 1.677~$\mu$m lines here we are limited to using the 1.534~$\mu$m / 1.644~$\mu$m ratio. From this ratio and the models we estimate ne $\sim$ 10,000 cm$^{-3}$. This is in line with typical jet densities \citep{Davis11}.

The mass outflow rate ($\dot{M}_{out}$) can be investigated by calculating the mass of both H$_{2}$ and H (from the [Fe II] lines) in the jet and combining this with the jet velocity and the length of the jet over for which the mass was estimated. Using equations 1 and 2 of \cite{Davis11} the column density of H$_{2}$ and H are estimated at N$_{H2}$ = 3.6 $\times$ 10$^{21}$ m$^{-2}$ and N$_{H}$ =1.5 $\times$ 10$^{22}$ m$^{-2}$. The fluxes of the H$_{2}$ 1-0S(1) and [Fe II]~1.644~$\mu$m lines were used here and were corrected for the jet extinction (A$_{v}$ = 9~mag). For the jet velocity the same approach as adopted in \cite{GLopez2013} is used here. That is the line width is taken as a lower limit to the jet velocity. The lower limits to the jet velocity are 100~km~s$^{-1}$ and 150~km~s$^{-1}$ for the H$_{2}$ and [Fe II] line respectively. For $\dot{M}_{out}$ it is found that $\dot{M}_{out[H2]}$ = 3.8~$\sim$ 10$^{-10}$~\Msun~yr$^{-1}$ and $\dot{M}_{out[FeII]}$ = 1.2~$\times$ 10$^{-9}$~\Msun~yr$^{-1}$. For $\dot{M}_{out[FeII]}$ a Fe / H solar abundance was assumed. Using the ratio of the [FeII]~1.257~$\mu$m and [PII]~1.188~$\mu$m lines it is found that 88$\%$~$\pm$~14 $\%$ of the Fe is still in dust grains in the jet \citep{GLopez2010}. Taking this value of the dust depletion into account places $\dot{M}_{out[FeII]}$ at $\sim$ 1 $\times$ 10$^{-8}$~\Msun~yr$^{-1}$. Overall our results put the jet efficiency ($\dot{M}_{out}$ / $\dot{M}_{acc}$) within the limits of magneto-centrifugal jet launching models \citep{Frank14}. Also note that as the mass outflow rate is higher in the [Fe II] than H$_{2}$ component of the jet, our results support the argument of \cite{GLopez2013}, that most of the outflow material is transported in the component traced by the atomic rather than molecular lines.

\begin{figure}
\begin{center}
   \includegraphics[width=9cm, trim= 0cm 1.5cm 1cm 2.5cm, clip=true]{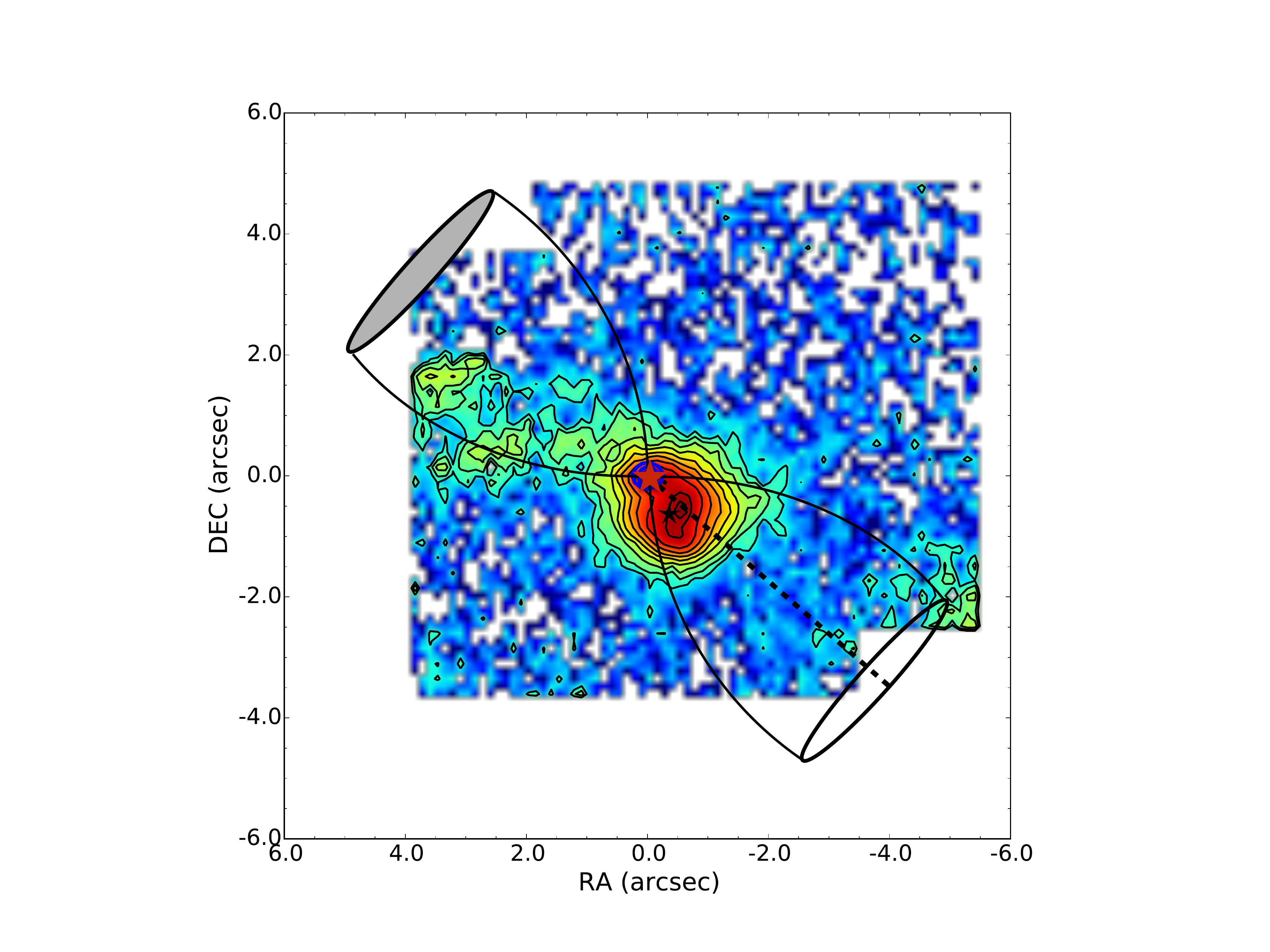}
     \caption{Colour map of the full H$_{2}$ 1-0 S(1) emission. The colours are shown to highlight the difference intensity between the H$_{2}$ jet emission along a PA of 230$^{\circ}$ (dashed line) and the extended emission which likely delineates a cavity. A cavity with opening angle 40$^{\circ}$ is overlain here and well represents the extended fainter emission.}
  \label{cavity}
  \end{center}     
\end{figure}

\section{Discussion and conclusions}

High angular resolution observations of jets from low mass YSOs have provided critical constraints to jet launching models in the form of collimation, rotation, and structural studies for example \citep{Frank14}. The difficulty with obtaining such high angular resolution observations, of jets at the lower end of the mass spectrum, has meant that the question of whether there are differences in how BD jets and stellar jets are launched has not yet been well investigated \citep{Whelan12, GLopez2013}. To date, discussion of BD jet launching has focused on a comparison between the kinematical and morphological properties of BD and YSO jets (e.g. asymmetries, multiple velocity components, molecular components) and on measurements of the jet efficiency. High angular resolution is not necessary to measure the jet efficiency and magneto-centrifugal jet launching models set an upper limit of 30$\%$ for $\dot{M}_{out}$ / $\dot{M}_{acc}$ \citep{Cabrit09}. Thus, studies of $\dot{M}_{out}$ / $\dot{M}_{acc}$ have been used as a first test of jet launching at the lowest masses \citep{Whelan14a}.

The detection here of a molecular hydrogen jet and cavity adds to the list of ways in which jets at the lowest masses are found to be analogous to stellar jets. Molecular hydrogen emission lines (MHELs) have long been associated with stellar jets with two jet components detected. Both a molecular component to a collimated FEL jet and a component which traces excitation along the walls of a edge-brightened cavity are observed \citep{Davis02}. For the latter component, it is theorised that the jet is surrounded by a wide-angled wind which carves out a cavity in the ambient medium and that the shocked H$_{2}$ emission is a result of the interaction between the wind and the ambient gas \citep{Frank14}. \cite{Davis02} give the wide-angled wind opening angles of a number of Class 0/I sources. Values range between 25$^{\circ}$ and 100$^{\circ}$ and there is a trend of the opening angle increasing with age. In Figure \ref{cavity}, the H$_{2}$ cavity emission is shown in green over-plotted on the H$_{2}$ jet emission (red). The extended H$_{2}$ emission from ISO-Oph~200 is fitted with a bipolar cavity of opening angle of 40$^{\circ}$, putting our source within the range of the Class I sources of \cite{Davis02}. The fact that the extended H$_{2}$ emission has different geometry and kinematics to the blue-shifted jet as traced by both H$_{2}$ and [Fe II] supports the argument that what we are seeing is a wind-swept cavity. A slow wide-angled wind is a natural by-product of a magneto-hydrodynamic (MHD) disk wind and its detection here is further evidence that MHD jets occur at the lowest masses \citep{Frank14}.

Turning now to estimates of jet efficiencies for proto-BDs, it would be reasonable to expect $\dot{M}_{out}$ in proto-BDs to be higher than for Class II BDs, as seen for low mass YSOs \citep{Frank14}, and for the jet efficiencies to remain within the limits of leading jet launching models. \cite{Riaz17} presented some evidence that $\dot{M}_{out}$ in proto-BDs is higher than in Class II BDs but very few sources have been properly investigated to date. The first studies of $\dot{M}_{out}$ / $\dot{M}_{acc}$ in BDs concentrated on Class II sources and values were found to be $>$ 1 which is inconsistent with models. Improvements in data and methods led to a better constrained estimate for one Class II BD ISO-Cha I 217 \citep{Whelan14a}. Observations of proto-BD jets have to date been confined to optical data which has led to problems with estimates of jet efficiencies. Firstly, the jet and source extinction is generally not well known and secondly $\dot{M}_{acc}$ has been under-estimated due to poor sampling of the accretion zone by the accretion tracers. The efficiency study presented here demonstrates the importance of NIR observations for investigating $\dot{M}_{out}$ / $\dot{M}_{acc}$ in BD jets. The [Fe II] and H$_{2}$ lines allow both the jet and source extinction to be well studied, the dust depletion in the jet to be investigated, and obscuration effects are limited by the use of the NIR HI lines. These improvements have meant that this is the second case where efficiency has been found to be within accepted limits. Also note that $\dot{M}_{out}$ in ISO-Oph 200 ($\sim$ 10$^{-9}$) is higher than in ISO-ChaI 217 ($\sim$ 10$^{-11}$) as would be expected.

In conclusion we argue that NIR studies of proto-BD jets are an important tool for investigating jet launching in the BD mass regime. Evidence of outflow components such as wide-angled winds can be looked for, the source and jet extinction can be investigated and measurements of the jet efficiency made. Furthermore, it will be possible in the near future to carry out observations of the jet collimation and structure at high angular resolution with JWST / NIRSPEC, which can provide additional constraints to jet launching models.


{}

\appendix
{
\section{Emission lines fluxes and excitation diagram}

\begin{table}
\centering
\begin{tabular}{ccc}        
\hline \hline 
&$\lambda_{vac}$ ($\mu$m)    &Flux ($\times$ 10$^{-16}$~erg~cm$^{-2}$~s$^{-1}$)
\\
\hline
H$_{2}$ 3-1S(7) &1.13 &8.5~$\pm$~0.6
\\
$[PII]$            &1.188 &3.7~$\pm$~0.6
\\
$[Fe II]$ &1.257    &26.3~$\pm$~0.7
\\ 
Pa$\beta$ &1.2822   &120.0~$\pm$~0.7
\\
$[Fe II]$ &1.279    &18.6~$\pm$~0.7
\\ 
$[Fe II]$ &1.295    &7.9~$\pm$~0.7
\\ 
$[Fe II]$ &1.321    &9.2~$\pm$~0.7
\\ 
$[Fe II]$ &1.534    &10.6~$\pm$~1.7
\\ 
Br12 &1.6412    &13.6~$\pm$~1.9
\\
$[Fe II]$ &1.644   &47.6~$\pm$~1.9
\\
H$_{2}$ 1-0S(3)  & 1.9576 & 33.7~$\pm$~1.5
\\
H$_{2}$ 1-0S(2) &2.0338  &33.8~$\pm$~1.5
\\
H$_{2}$ 1-0S(1) &2.1218  &72.2~$\pm$~1.6 
\\
Br$\gamma$ &2.1661    &305.0~$\pm$~1.6
\\
H$_{2}$ 1-0S(0) &2.2235  &21.5~$\pm$~1.7
\\
H$_{2}$ 2-1S(1) &2.2477  &7.3~$\pm$~1.7 
\\
H$_{2}$ 1-0Q(1) &2.4066  &72.4~$\pm$~3.6 
\\
H$_{2}$ 1-0Q(2) &2.4134  &33.1~$\pm$~3.6 
\\
H$_{2}$ 1-0Q(3) &2.4237  &74.4~$\pm$~3.6 
\\
\hline 
\end{tabular}
\caption{Table of detected lines. The fluxes represent the source and jet emission and have not been extinction corrected.}
\label{table}
\end{table}

}

\begin{figure}
\begin{center}
   \includegraphics[width=8cm, trim= 0cm 0cm 0cm 0cm, clip=true]{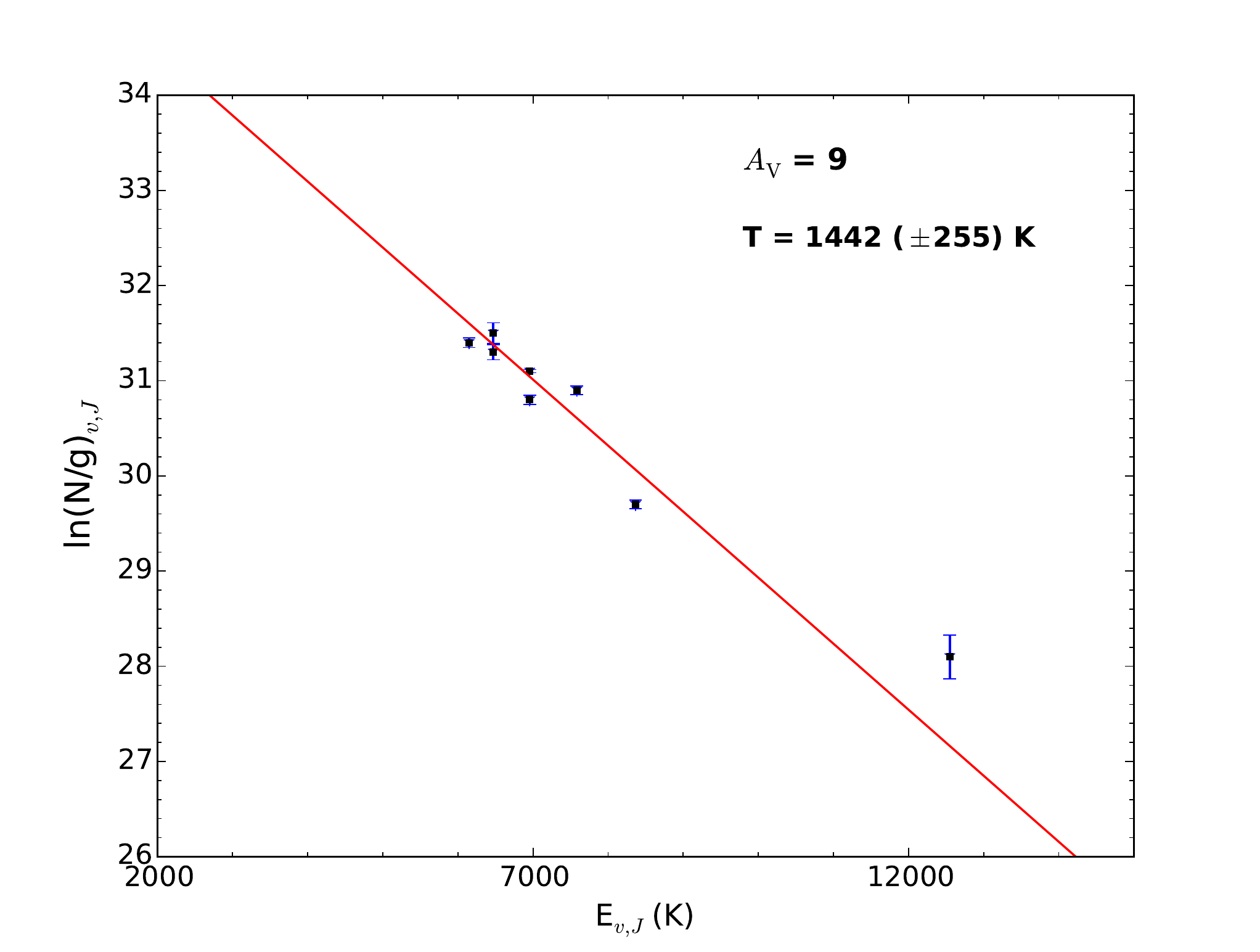}
     \caption{Colour map of the full H$_{2}$ 1-0 S(1) emission. The colours are shown to highlight the difference intensity between the H$_{2}$ jet emission along a PA of 230$^{\circ}$ (dashed line) and the extended emission which likely delineates a cavity. A cavity with opening angle 40$^{\circ}$ is overlain here and well represents the extended fainter emission.}
  \label{rot}
  \end{center}     
\end{figure}


\begin{thebibliography}{}

\bibitem[Alcal{\'a} et al.(2017)]{Alcala17} Alcal{\'a}, J.~M., Manara, C.~F., Natta, A., et al.\ 2017, \aap, 600, A20 

\bibitem[Antoniucci et al.(2014)]{Ant2014} Antoniucci, S., La Camera, A., Nisini, B., et al.\ 2014, \aap, 566, A129

\bibitem[Beck(2007)]{Beck07} Beck, T.~L.\ 2007, \aj, 133, 1673

\bibitem[Baraffe et al.(2017)]{Baraffe17} Baraffe, I., Elbakyan, V.~G., Vorobyov, E.~I., \& Chabrier, G.\ 2017, \aap, 597, A19  

\bibitem[Cabrit(2009)]{Cabrit09} Cabrit, S.\ 2009, Astrophysics and Space Science Proceedings, 13, 247 

\bibitem[Calvet et al.(2004)]{Calvet04} Calvet, N., Muzerolle, J., Brice{\~n}o, C., et al.\ 2004, \aj, 128, 1294

\bibitem[Caratti o Garatti et al.(2006)]{Caratti2006} Caratti o Garatti, A., Giannini, T., Nisini, B., \& Lorenzetti, D.\ 2006, \aap, 449, 1077 

\bibitem[Cardelli et al.(1989)]{Cardelli89} Cardelli, J.~A., Clayton, G.~C., \& Mathis, J.~S.\ 1989, \apj, 345, 245

\bibitem[Cutri et al.(2003)]{Cutri03} Cutri, R.~M., Skrutskie, M.~F., van Dyk, S., et al.\ 2003, VizieR Online Data Catalog, 2246,

\bibitem[Davis et al.(2002)]{Davis02} Davis, C.~J., Stern, L., Ray, T.~P., \& Chrysostomou, A.\ 2002, \aap, 382, 1021 

\bibitem[Davis et al.(2006)]{2006ApJ...639..969D} Davis, C.~J., Nisini, B., Takami, M., et al.\ 2006, \apj, 639, 969 

\bibitem[Davis et al.(2011)]{Davis11} Davis, C.~J., Cervantes, B., Nisini, B., et al.\ 2011, \aap, 528, A3

\bibitem[Evans et al.(2009)]{Evans2009} Evans, N.~J., II, Dunham, M.~M., J{\o}rgensen, J.~K., et al.\ 2009, \apjs, 181, 321-350

\bibitem[Frank et al.(2014)]{Frank14} Frank, A., Ray, T.~P., 
Cabrit, S., et al.\ 2014, Protostars and Planets VI, 451

\bibitem[Garcia Lopez et al.(2010)]{GLopez2010} Garcia Lopez, R., Nisini, B., Eisl{\"o}ffel, J., et al.\ 2010, \aap, 511, A5 

\bibitem[Garcia Lopez et al.(2013)]{GLopez2013} Garcia Lopez, R., Caratti o Garatti, A., Weigelt, G., Nisini, B., \& Antoniucci, S.\ 2013, \aap, 552, L2

\bibitem[Giannini et al.(2013)]{Giannini2013} Giannini, T., Nisini, B., Antoniucci, S., et al.\ 2013, \apj, 778, 71

\bibitem[Giannini et al.(2015)]{Giannini2015} Giannini, T., Antoniucci, S., Nisini, B., et al.\ 2015, \apj, 798, 33 

\bibitem[Gillessen et al.(2005)]{Gillessen05} Gillessen, S., Davies, R., Kissler-Patig, M., et al.\ 2005, The Messenger, 120, 26 

\bibitem[Gredel(1994)]{Gredel94} Gredel, R.\ 1994, \aap, 292, 580 

\bibitem[Nisini et al.(2005)]{Nisini05} Nisini, B., Bacciotti, F., Giannini, T., et al.\ 2005, \aap, 441, 159 

 \bibitem[Pecchioli et al.(2016)]{Pecc2016} Pecchioli, T., Sanna, N., Massi, F., \& Oliva, E.\ 2016, \pasp, 128, 073001 

\bibitem[Nisini et al.(2016)]{Nisini16} Nisini, B., Giannini, T., Antoniucci, S., et al.\ 2016, \aap, 595, A76 
 
\bibitem[Riaz \& Whelan(2015)]{Riaz15} Riaz, B., \& Whelan, E.~T.\ 2015, \apjl, 815, L31

\bibitem[Riaz et al.(2016)]{Riaz2016} Riaz, B., Vorobyov, E., Harsono, D., et al.\ 2016, \apj, 831, 189 

\bibitem[Riaz et al.(2017)]{Riaz17} Riaz, B., Brice{\~n}o, C., Whelan, E.~T., \& Heathcote, S.\ 2017, \apj, 844, 47

\bibitem[Takami et al.(2006)]{Takami06} Takami, M., Chrysostomou, A., Ray, T.~P., et al.\ 2006, \apj, 641, 357 

\bibitem[Todd \& Ramsay Howat(2006)]{Todd06} Todd, S.~P., \& Ramsay Howat, S.~K.\ 2006, \mnras, 367, 238 

\bibitem[Whelan et al.(2004)]{Whelan2004} Whelan, E.~T., Ray, T.~P., \& Davis, C.~J.\ 2004, \aap, 417, 247


\bibitem[Whelan et al.(2012)]{Whelan12} Whelan, E.~T., Ray, T.~P., Comeron, F., Bacciotti, F., \& Kavanagh, P.~J.\ 2012, \apj, 761, 120 

\bibitem[Whelan(2014a)]{Whelan14a} Whelan, E.~T.\ 2014a, Astronomische Nachrichten, 335, 537 

\bibitem[Whelan et al.(2014b)]{Whelan14b} Whelan, E.~T., Bonito, R., Antoniucci, S., et al.\ 2014b, \aap, 565, A80 

\end{thebibliography}
\end{document}